\documentclass[10pt,american,preprint, aps, nofootinbib]{revtex4-1}
\usepackage[T1]{fontenc}
\usepackage[latin9]{inputenc}
\usepackage{babel}
\usepackage{mathrsfs}
\usepackage{amsmath}
\usepackage{amssymb}
\usepackage{graphicx}
\usepackage[pdfusetitle,
 bookmarks=true,bookmarksnumbered=false,bookmarksopen=false,
 breaklinks=false,pdfborder={0 0 1},backref=false,colorlinks=false]
 {hyperref}

\makeatletter
\usepackage{etoolbox}
\usepackage{breqn}

\makeatletter
\let\cat@comma@active\@empty
\makeatother

\newcommand{\breqnoverloadothers}
{% 
    \renewenvironment{equation}{\ignorespaces\begin{dmath}}{\end{dmath}\ignorespacesafterend}%
    \renewenvironment{equation*}{\ignorespaces\begin{dmath*}}{\end{dmath*}\ignorespacesafterend}%
    \renewenvironment{multline}{\ignorespaces\begin{dmath}}{\end{dmath}\ignorespacesafterend}%
    \renewenvironment{multline*}{\ignorespaces\begin{dmath*}}{\end{dmath*}\ignorespacesafterend}%

}

\newcommand\breqnundefineothers
{%  
    \renewenvironment{equation}{}{}%
    \renewenvironment{equation*}{}{}%
    \renewenvironment{multline*}{}{}%

}

\AtBeginEnvironment{dmath}{\breqnundefineothers}
\AtBeginEnvironment{dmath*}{\breqnundefineothers}

\AtBeginEnvironment{dgroup}{\def\breqnundefineothers{}\breqnoverloadothers}
\AtBeginEnvironment{dgroup*}{\def\breqnundefineothers{}\breqnoverloadothers}

\newcommand\brwrap[3]{%
  \setbox0=\hbox{$#2$}
  \left#1\vbox to \the\ht0{\hbox to 0pt{}}\right.\kern-.2em
  \begingroup #2\endgroup\kern-.15em
  \left.\vbox to \the\ht0{\hbox to 0pt{}}\right#3
}

\ifdefined\showcaptionsetup
 \PassOptionsToPackage{caption=false}{subfig}
\fi
\usepackage{subfig}
\makeatother

\begin{document}
\title{Breakdown of Semiclassical Gravity in Four-Dimensional Black Hole
Evaporation}
\author{David A. Lowe}
\affiliation{Department of Physics, Brown University, Providence, RI 02912, USA}
\author{Larus Thorlacius}
\affiliation{Science Institute, University of Iceland, Dunhaga 3, 107 Reykjavik,
Iceland }
\begin{abstract}
We study black hole formation and evaporation in a four-dimensional
semiclassical model that preserves diffeomorphism invariance and reproduces
the one-loop trace anomaly. Solving the quantum-corrected Einstein
equations for the collapse of a spherically symmetric null shell,
we follow the formation and evaporation of a black hole with back-reaction
included. The semiclassical solutions develop a spacelike \textquotedblleft thunderbolt\textquotedblright{}
singularity that emerges after the apparent horizon has receded and
extends far from the black hole where the semiclassical curvature
is a priori expected to be parametrically small. This behavior arises
from a nonlinear instability of the higher-derivative semiclassical
equations and is generic in models with anomaly-induced quantum corrections.
The thunderbolt signals a breakdown of semiclassical effective field
theory over macroscopic distances and undermines the standard formulation
of the black hole information paradox.
\end{abstract}
\maketitle

\section{Introduction\label{sec:Introduction}}

The black hole information paradox \citep{Hawking:1976ra} remains
one of the central problems in quantum gravity. The standard argument
for information loss is made in the context of semiclassical gravity,
a low-energy effective field theory of gravity coupled to matter,
assumed to have solutions that describe the formation and subsequent
evaporation of macroscopic black holes in asymptotically flat spacetime.
Much of this intuition about semiclassical black hole evaporation
has been shaped by lower-dimensional toy models, most notably two-dimensional
dilaton gravity, where calculations are tractable and explicit analytic
results can be obtained \citep{Callan:1992rs,Russo:1992ax,Lowe:1992ed}.
The situation is less clear in four spacetime dimensions. While Hawking\textquoteright s
original calculation demonstrated that black holes radiate \citep{HAWKING_1974,Hawking1975},
incorporating the back-reaction of this radiation on the spacetime
geometry has proven technically challenging. The nonlocal structure
of the effective action induced by quantum matter fields, together
with the complexity of the four-dimensional semiclassical Einstein
equations, has held back progress toward a definitive description
of four-dimensional black hole evaporation. 

In the present paper, we revisit the back-reaction problem using a
four-dimensional semiclassical model that retains full diffeomorphism
invariance and exactly reproduces the one-loop trace anomaly. The
model is based on the Riegert anomaly-induced action \citep{RIEGERT198456}
coupled to Einstein gravity, which can be viewed as a controlled truncation
of the full one-loop effective action. While not complete, this framework
captures essential quantum effects associated with conformal matter
fields and provides a concrete setting in which semiclassical back-reaction
can be studied dynamically. Earlier literature on this effective action,
and some generalizations of it, includes \citep{Balbinot:1998yh,Balbinot:1999ri,Balbinot:aa,Mottola:2006ew,Mottola_2011,Fabris_2012,Mottola:2016aa,Mottola:2023jlo,Mottola_2023,Mottola:2025fhl,Lowe:2025bxl,Liu:2025xfu,Lowe:2025tik}.
In particular, the recent works \citep{Lowe:2025bxl,Liu:2025xfu}
focussed on Hawking emission in a static four-dimensional Schwarzschild
background, where an Unruh-like solution \citep{Unruh:1976aa} emerged
as essentially the unique solution that satisfied regularity conditions
on the horizon and near spacelike infinity, while \citep{Lowe:2025tik}
considered the time-dependent problem of Hawking emission from a black
hole formed in gravitational collapse, working in the approximation
of a fixed spacetime background. This allowed the long-standing issue
of the onset of Hawking radiation to be resolved in a four-dimensional
approach. 

In the following, we go beyond the fixed background spacetime approximation
and study the full set of semiclassical field equations with back-reaction
included. We consider spherically symmetric configurations and numerically
solve the resulting higher-derivative equations of motion using a
characteristic formulation in double-null coordinates. This allows
us to track the formation and evolution of an apparent horizon and
thereby follow the evaporation process. A central result of this paper
is the appearance of a spacelike thunderbolt singularity that develops
after the apparent horizon has receded and the black hole has effectively
evaporated, as originally suggested by Hawking and Stewart \citep{Hawking:1992ti}
in the context of two-dimensional dilaton gravity.\footnote{Subsequent work, however, showed that this feature does not actually
appear in the two-dimensional models \citep{Lowe:1992ed,Piran:1993tq}.} Strikingly, this singularity extends into the region outside the
black hole, where semiclassical curvature would \emph{a priori} be
expected to be small. We provide both numerical evidence and analytic
arguments supporting the generic nature of this phenomenon. The numerical
evidence points to the region near past null infinity being smooth,
ruling out a naive ghost-mode instability. Instead the thunderbolt,
which appears to extend to spacelike infinity, is a direct consequence
of the back-reaction of the outgoing radiation on the spacetime geometry.
Our results establish a sharp qualitative distinction between two-dimensional
semiclassical gravity, which suffers from no such instability, and
four-dimensional semiclassical gravity where the instability follows
from the fourth-order structure of the field equations.

If the emergence of the thunderbolt singularity were an observable
physical effect, the evaporation of a single primordial black hole
would have had catastrophic consequences on a cosmological scale in
our own universe. Instead, it signals a breakdown of the semiclassical
description of black hole evaporation. A key question going forward
is then whether, and how, the description can be improved to give
a more plausible view of gravitational collapse in four spacetime
dimensions with an asymptotically flat exterior geometry. For instance,
is it possible to eliminate the thunderbolt by fine-tuning the initial
data while leaving the field equations unchanged? While we have not
been able to rule out this possibility, we find it more likely that
the semiclassical model itself needs to be modified and that requiring
the absence of thunderbolt singularities will place novel constraints
on effective field theory for gravity in four spacetime dimensions.

The organization of the paper is as follows. In Section \eqref{sec:Basic-Setup}
we describe the semiclassical model, the metric ansatz, and the numerical
method. In Section \eqref{sec:Numerical-Results} we analyze black
hole evaporation and the behavior of curvature and stress-energy.
Section \eqref{sec:Thunderbolt-Singularity} presents evidence for
the thunderbolt singularity and discusses its physical origin. We
conclude in Section \eqref{sec:Conclusions} with implications for
semiclassical gravity and the black hole information paradox.

\section{Basic Setup\label{sec:Basic-Setup}}

We begin with the single scalar Riegert action coupled to Einstein
gravity in four-dimensional spacetime \citep{RIEGERT198456}. The
action takes the form\begin{dmath}
\begin{equation}
S=\int d^{4}x\,(-g)^{1/2}\left[\frac{1}{16\pi}R+\frac{1}{192\pi^{2}}(c-\tfrac{2}{3}b)R^{2}-\tfrac{b}{2}\nabla^{2}\phi\nabla^{2}\phi-\tfrac{b}{3}R(\nabla\phi)^{2}+bR^{ab}\nabla_{a}\phi\nabla_{b}\phi+\frac{\phi}{8\pi}\left((a+b)C^{2}+\frac{2b}{3}\left(R^{2}-3R_{ab}R^{ab}-\nabla^{2}R\right)\right)\right],\label{eq:action}
\end{equation}
\end{dmath} where we use the so-called $(+++)$ conventions of \citep{misner}
and define $C^{2}$ as the square of the Weyl tensor. The scalar field
$\phi$ localizes the original non-local Riegert action, which depends
only on the dynamical metric $g_{\mu\nu}$. The trace anomaly depends
on the three coefficients $a,b$ and $c$, which depend on the matter
content of the theory. We adopt a semiclassical limit, with a large
but finite number of matter fields $N$, so that matter quantum effects
dominate over graviton loops. 

Often the anomaly coefficient $c$ is not considered, because it can
be defined away by a shift in a local counter-term (i.e. $R^{2}$),
however we will find it convenient to retain it. Note that the $b$
coefficient is negative for matter fields of low spin $s\leq1$ \citep{Duff:1993wm}.
This contributes a negative sign to the energy flux of outgoing Hawking
radiation at future infinity \citep{Lowe:2025bxl}, but a positive
overall outgoing energy flux can be arranged by introducing a second
auxiliary scalar field that couples to the square of the Weyl tensor,
as discussed in \citep{Liu:2025xfu}. The terms involving the second
scalar field add an unwanted layer of complexity to numerical computations
and thus we work with the unadorned Riegert model and simply set $b>0$
by hand to obtain a positive outgoing energy flux. 

We make a metric ansatz,
\begin{equation}
ds^{2}=-e^{2\rho}dudv+r^{2}d\Omega^{2},\label{eq:metric}
\end{equation}
allowing a time-dependent and spherically symmetric dependence via
$\rho=\rho(u,v)$ and $r=r(u,v)$. We substitute into the action and
directly derive the equations of motion for the dynamical fields $\rho,r$
and $\phi$.

The resulting equations of motion are complex and not particularly
illuminating to write out explicitly.  In practice, we have used the
Mathematica package xAct \citep{Martin-Garcia:2007bqa,Martin-Garcia:2008ysv}
to evaluate the action using the form of the metric \eqref{eq:metric}.
We then use the Mathematica Variational Methods package to derive
the equations of motion. These take the form of 4th order nonlinear
partial differential equations for the variables $\rho(u,v),\,r(u,v)$
and $\phi(u,v)$. We choose to integrate these equations using the
method of characteristics, essentially treating the null coordinate
$u$ as a time-direction, and for each fixed value of $u$, integrating
along the $v$-direction. In essence, each time step is reduced to
solving an ordinary differential equation in the $v-$direction, followed
by another ordinary differential equation step in the $u$-direction.
The method of characteristics code uses 4th order Runge-Kutta methods
to perform the ordinary differential equation solve on a regular rectangular
lattice in the $(u,v)$ plane. This is encoded in MATLAB, and we use
a MATHEMATICA-to-MATLAB conversion package to translate the equations
of motion into the code. In this formulation convergence can be achieved
by ensuring the step-sizes in each direction are sufficiently small
relative to the derivative terms. Given the complexity of the equations,
this is best tested by multiple runs with varying grid sizes.

To define the boundary data, we restrict the numerical solution to
a coordinate patch to the future of an ingoing null shock wave. We
are free to take the classical metric in this region to be the Kruskal
solution
\begin{equation}
ds^{2}=-\frac{32M^{3}}{r}e^{-\frac{r}{2M}}dUdV+r^{2}d\Omega^{2},\label{eq:lineelement}
\end{equation}
where 
\begin{equation}
r(U,V)=2M\left(1+W\left(-UV/e\right)\right),\label{eq:rofuv}
\end{equation}
and $W$ is the Lambert function. In practice the choice of Kruskal
coordinates is difficult to handle numerically, so we perform a stretch
of the coordinates, namely we define new coordinates $u$ and $v$
such that
\begin{align}
U & =A-e^{-u/4},\label{eq:coords}\\
V & =e^{v/4},\nonumber 
\end{align}
introducing a new parameter $A$. For $A=0$ one recovers null coordinates
related to the usual tortoise coordinate $r_{*}$ which diverges on
the horizon. Here we require the coordinate patch covers points inside
the apparent horizon, which will be true if $A>0$. The precise value
of $A$ can be adjusted so as to ensure the numerics samples many
points in the neighborhood of the apparent horizon. 

We work on a rectangular coordinate patch,
\begin{equation}
v_{min}<v<v_{max}\,,\qquad u_{min}<u<u_{max}\,,\label{eq:patch}
\end{equation}
where $v=v_{min}$ corresponds to the ingoing null shock wave, $v=v_{max}$
and $u=u_{min}<0$ are surrogates for future and past null infinity,
respectively, and $u=u_{max}>0$ is chosen large enough to ensure
the coordinate patch extends into the future trapped region inside
the apparent horizon of the black hole.

The method of characteristics requires initial data for the fields
and their first derivatives along two of the null edges of the coordinate
patch. Let us denote by $r_{0}(u,v)$ and $\rho_{0}(u,v)$ the values
of these variables on the classical Kruskal solution \eqref{eq:lineelement}
after transforming to the coordinates \eqref{eq:coords}. Initial
data is then formulated by requiring
\begin{equation}
\begin{cases}
\phi=1,\,\partial_{u}\phi=0\\
r=r_{0},\,\partial_{u}r=\partial_{u}r_{0} & \quad\textrm{{for}}\quad u=u_{min},\,v>v_{min}\,,\\
\rho=\rho_{0},\,\partial_{u}\rho=\partial_{u}\rho_{0}
\end{cases}\label{eq:initialdata}
\end{equation}
and likewise along null shockwave we take
\begin{equation}
\begin{cases}
\phi=1,\,\partial_{v}\phi=0\\
r=r_{0},\,\partial_{v}r=\partial_{v}r_{0} & \quad\textrm{{for}}\quad u_{max}\geq u\geq u_{min},\,v=v_{min}\,.\\
\rho=\rho_{0},\,\partial_{v}\rho=\partial_{v}\rho_{0}
\end{cases}\label{eq:shockdata}
\end{equation}
In other words, along the initial null slice and along the ingoing
null shock, we take the metric to be that of a classical black hole
of mass $M$ and impose both Dirichlet and Neumann conditions on the
auxiliary scalar.\footnote{In order to avoid nontrivial corner conditions we must also impose
corresponding conditions at $v=v_{min}$ on $\partial^{2}_{u}X$ and
$\partial_{v}\partial^{2}_{u}X$ where $X=(\phi,r,\rho)$.} These conditions ensure no extra energy flux is incident from $\mathscr{I}^{-}$
and that likewise that no additional outgoing flux crosses the shockwave.
This no flux condition holds exactly for the conditions \eqref{eq:initialdata}
and \eqref{eq:shockdata}. Nevertheless, as we will see, the conditions
lead to a mild transient near the shock, which settles down within
a couple of light-crossing times.

To summarize, the first step is to integrate in the $v$-direction
at fixed $u=u_{min}$, starting at $v=v_{min}$, writing the equations
of motion,
\begin{equation}
\partial^{2}_{v}\partial^{2}_{u}X=f(X,\partial_{u}X,\partial_{v}X,\partial_{u}\partial_{v}X,\partial^{2}_{u}X,\partial_{v}\partial^{2}_{u}X),\label{eq:numericeqn}
\end{equation}
along a line of constant $u$, which solves for the ``acceleration''
$\partial^{2}_{u}X$ for general $v$ given the boundary data noted
above. Once one has that computed, it is straightforward to take a
time-step in the $u$ direction. The equations of motion can be written
in the form \eqref{eq:numericeqn} for generic values for the anomaly
coefficients $(a,b,c)$ and in the present work we will choose $a=b=c=10$.
If instead one were to choose $a=b$ but $c=0$, for example, then
the matrix of coefficients of the $\partial^{2}_{v}\partial^{2}_{u}X$
terms would be non-invertible, so some of the equations of motion
would have a lower order form and would require modifying the numerical
method on a case-by-case basis. While we work with $a=b=c$ for numerical
convenience, we have checked that the qualitative instability mechanism
does not rely on this tuning; its origin lies in the generic fourth-order
structure of the anomaly-induced equations. 

It should be noted that the type of fourth order PDEs we are solving
often contain unphysical spurious solutions, so it is helpful to review
some context before proceeding. A well-known example of a third order
equation of motion is that of a charged particle interacting with
an electromagnetic field. Such equations contain solutions corresponding
to self-acceleration which can lead to the particle acquiring infinite
energy. However, specifying finite energy on some initial value surface,
including both the energy of the particle and the electromagnetic
field, turns out to be sufficient to rule out such spurious solutions.
Our strategy here will likewise be to impose a version of smooth,
finite energy initial data on an initial value surface and study how
the solution develops.

In our earlier work \citep{Lowe:2025tik} we studied the simpler problem
where back-reaction was ignored, by simply solving the equation for
$\phi$ on a fixed background, following the strategy of the previous
paragraph. We found an instability did develop, but crucially this
ended up producing a bounded induced stress tensor which accounted
for the outgoing Hawking radiation at future null infinity, while
vanishing at (and near) past null infinity. The solution also exhibited
the near horizon onset of the outgoing Hawking radiation for the first
time from an intrinsically four-dimensional model. In the present
work, we include semiclassical back-reaction and find that in this
case smooth, finite energy initial data generically lead to unexpected
curvature singularities exterior to the black hole. \footnote{Note these considerations do not carry over straightforwardly to cosmological
solutions, where the constraint of smoothness on an initial slice
is far less restrictive.} 

\section{Numerical Results\label{sec:Numerical-Results}}

We numerically solve the PDEs for $M=2.2$, $a=b=c=10$ and take $(u_{min},v_{min})=(-50,1)$
to be the corner of our coordinate patch, with the shock running along
the line $v=v_{min}$ and $u=u_{min}$ being our substitute for $\mathscr{I}^{-}$.
With this value of $M$ the curvature on the horizon is still relatively
small in Planck units, while we can expect the black hole endpoint
to also be accessible before numerical errors build up. In order to
have the black hole lifetime $M^{3}/N$ larger than the light-crossing
time $M$, we require $M^{2}\gtrsim N$. A more precise statement
involves the rather large numerical factors present in \eqref{eq:action},
and with the above choices of parameters, we are in the desired regime.

\begin{figure}

\includegraphics[scale=0.3]{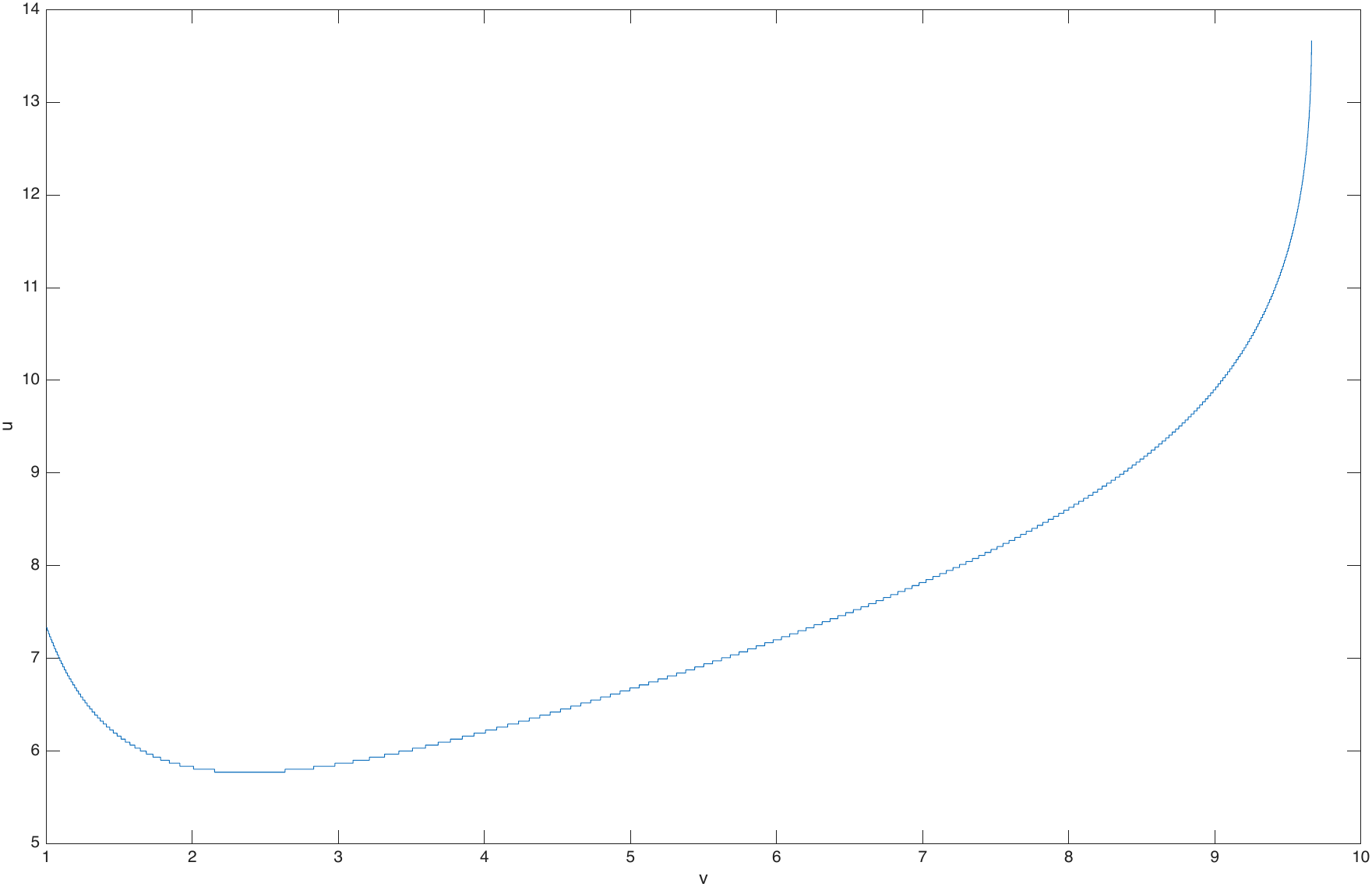}\includegraphics[scale=0.3]{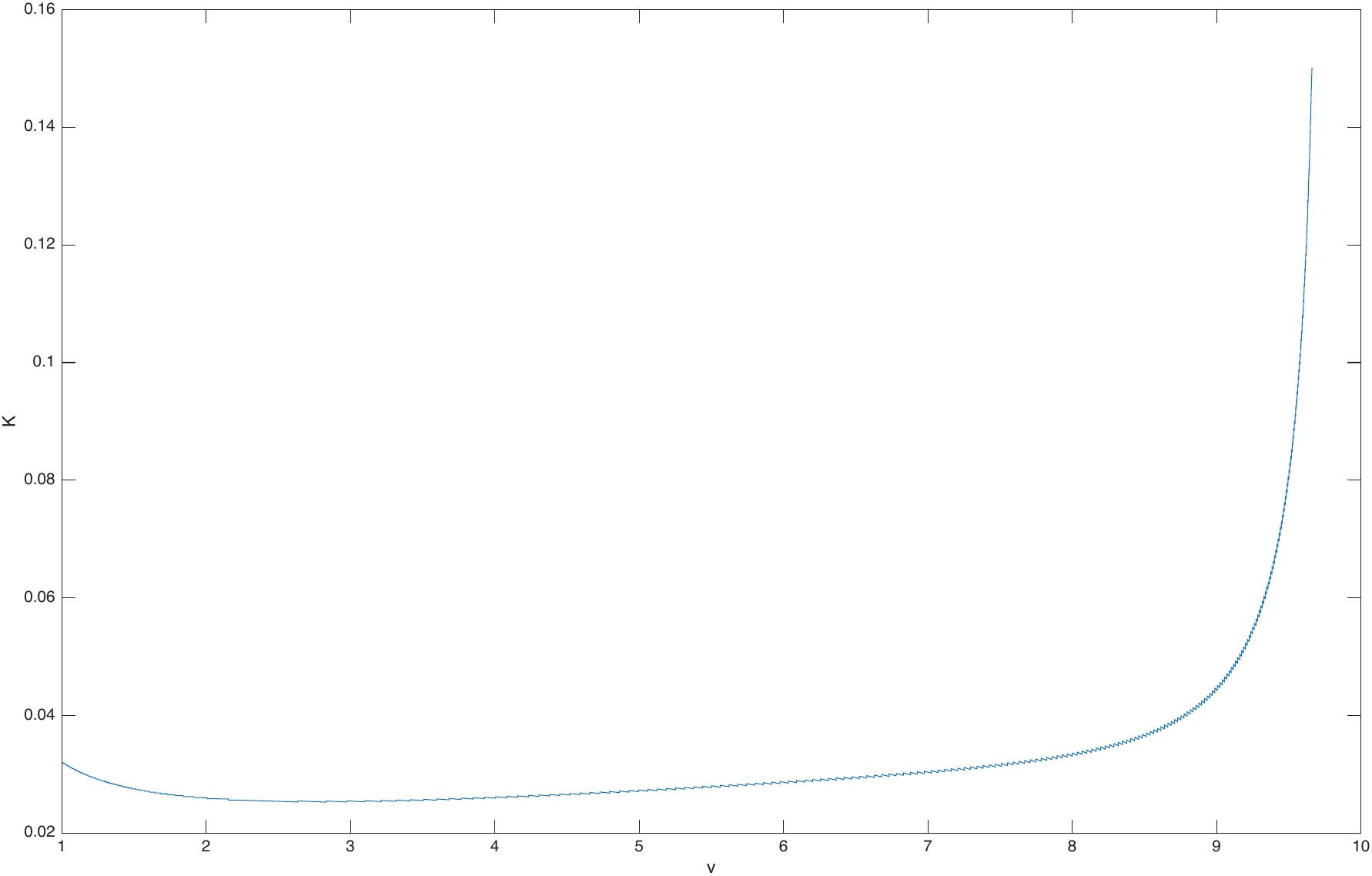}\caption{The position of the apparent horizon in the $u,v$ plane is shown
in the first panel. Following a brief transient after the shock at
$v=1$, the apparent horizon recedes, eventually leaving the coordinate
patch (with $A=0.16).$ In the second panel the Kretschmann scalar
is shown on the apparent horizon. This increases as expected for an
evaporating four-dimensional black hole.\label{fig:The-position-of}}

\end{figure}

An apparent horizon forms where $\partial_{v}r=0$. As the evaporation
proceeds, the apparent horizon recedes and the curvature on the horizon
(as measured by the Kretschmann scalar $K=R_{\mu\nu\lambda\rho}R^{\mu\nu\lambda\rho}$)
increases. The apparent horizon eventually leaves the coordinate patch
(the value of the parameter $A$ in \eqref{eq:coords} controls how
far beyond the classical horizon, $r=2M$, the patch extends). A simulation
is shown for the value $A=0.16$ performed on a $\left(4\times10^{3}\right)\times\left(20\times10^{3}\right)$
grid in figure \ref{fig:The-position-of}. We have verified point-wise
convergence in the patch displayed in the plots by increasing $|u_{min}|$
and by changing the number of grid points.

\begin{figure}

\includegraphics[scale=0.75]{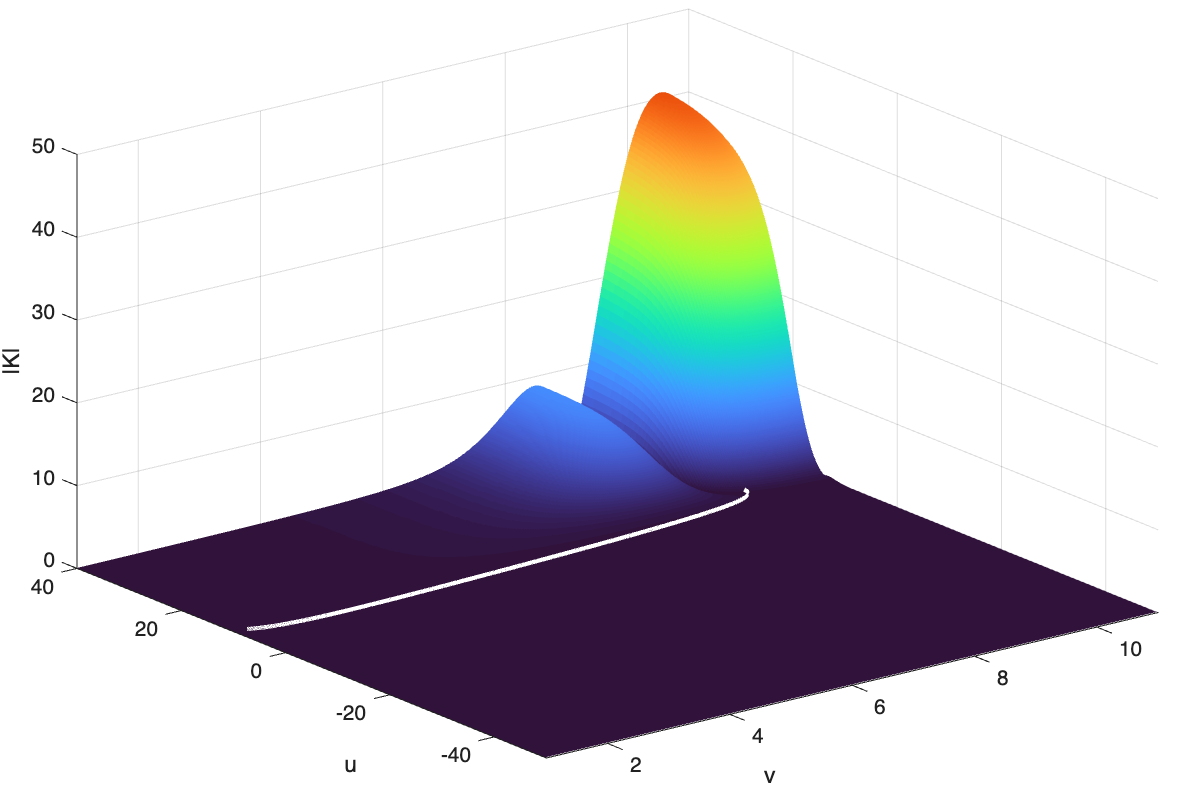}\caption{Surface plot of the absolute value of the Kretschmann curvature scalar
near the black hole endpoint. The white line shows the position of
the apparent horizon.\label{fig:Surface-plot-of}}

\end{figure}

As we will explore momentarily, the endpoint does not generate a Cauchy
horizon, where the curvature might fall off as one moves along a line
of constant $u$ toward $\mathscr{I}^{+}$. If
\begin{figure}

\includegraphics[scale=0.5]{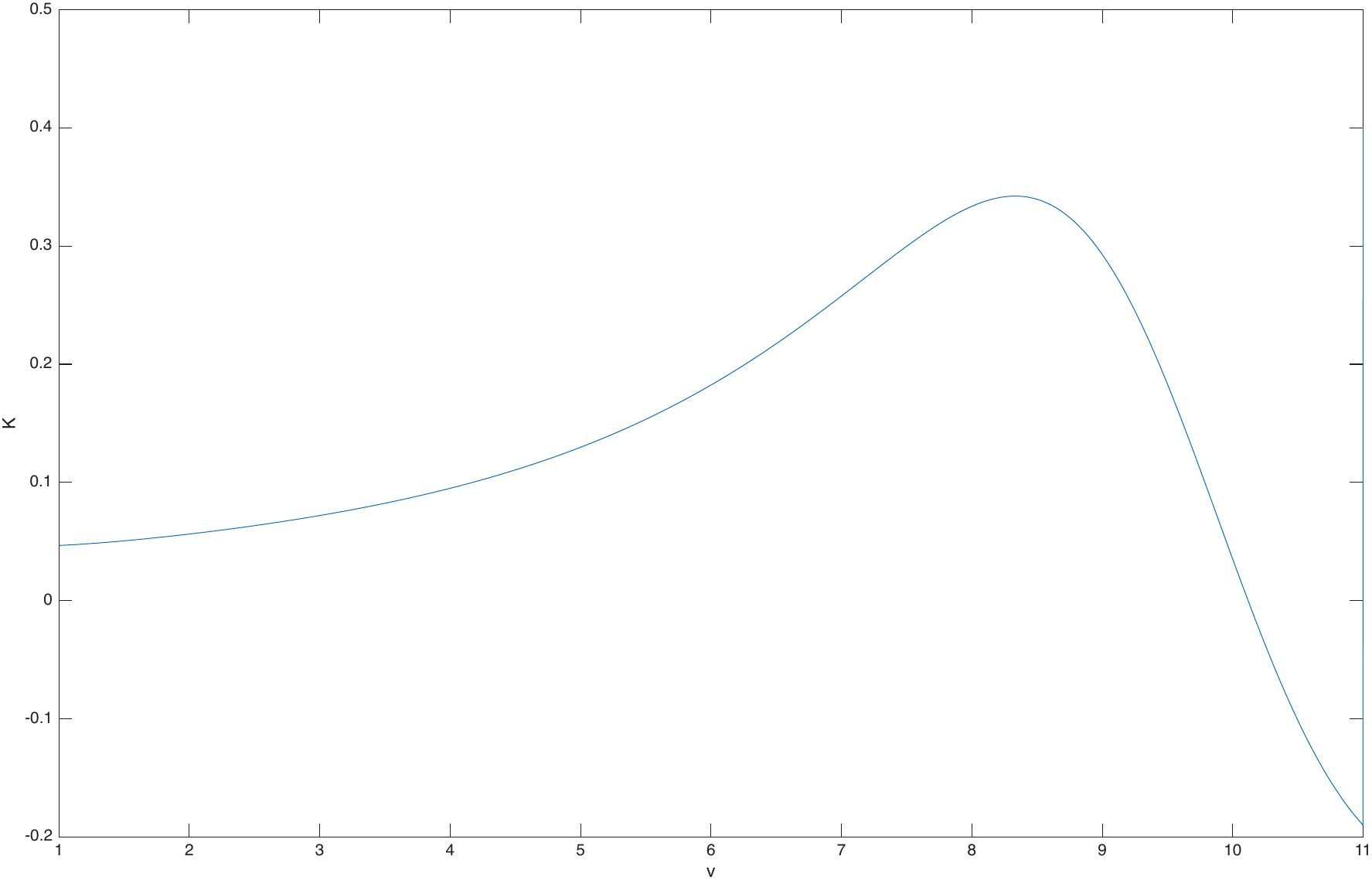}\caption{Kretschmann curvature scalar along a null ray emerging from the apparent
horizon.\label{fig:Curvature-along-a}}

\end{figure}
 we follow an outgoing null ray that starts inside the black hole
region, emerges through the apparent horizon, and proceeds outward,
we find the Kretschmann scalar behaves as shown in figure \ref{fig:Curvature-along-a}.
Initially the Kretschmann scalar increases, then decreases as the
null ray leaves the apparent horizon (at around $v=9$), but then
it passes through zero and becomes large and negative, indicating
the presence of the thunderbolt that we consider in more detail momentarily.
A surface plot of $|K|$ in the vicinity of the black hole endpoint
is shown in figure \ref{fig:Surface-plot-of}.

\begin{figure}

\includegraphics[scale=0.5]{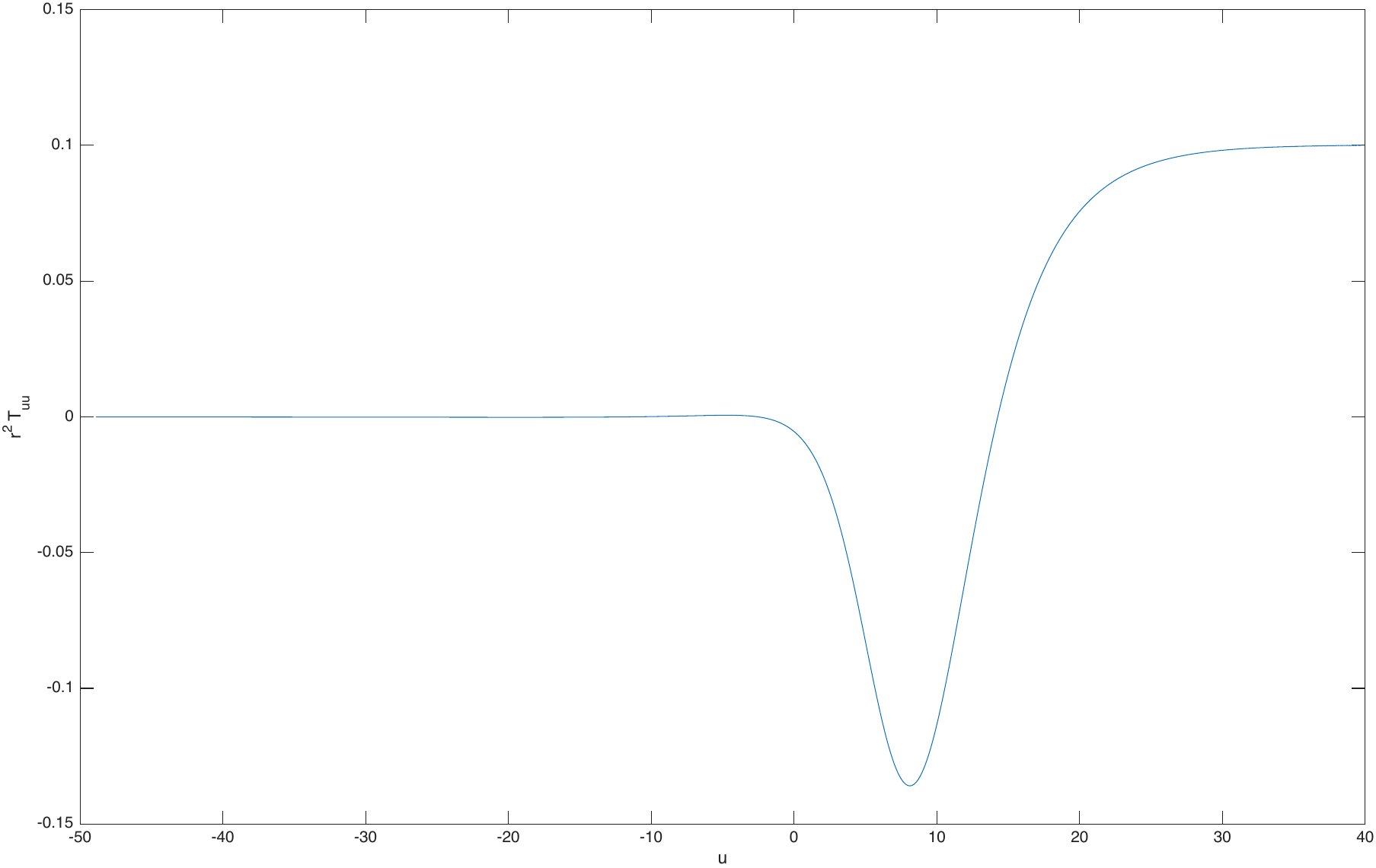}\caption{Outgoing energy flux as a function of $u$ at constant $v=30$, which
is outside the black hole but inside the thunderbolt. This shows the
expected vanishing result in the distant past, rising to a plateau.
The negative dip appears to be an artifact of not being able to choose
large $v$ due to the presence of the thunderbolt.\label{fig:Outgoing-stress-energy}}

\end{figure}

In the region where $u\geq0$, there is a range of $v$ outside the
black hole where the curvature is relatively small (before the thunderbolt
appears). Figure \ref{fig:Outgoing-stress-energy} shows the outgoing
energy flux $T_{uu}$ as a function of $u$ for constant $v$ in this
range. The result is broadly consistent with our previous result in
\citep{Lowe:2025tik} for the outgoing Hawking energy flux in a fixed
classical background spacetime, where a black hole is formed by gravitational
collapse but semiclassical back-reaction is neglected. In particular,
the energy flux vanishes at early retarded times and turns on as the
black hole is formed. 

It is interesting to consider whether $K$ in figure \ref{fig:The-position-of}
can be fitted to the expected $K\propto(t_{end}-t)^{-4/3}$ expected
for the adiabatic evolution of a large black hole emitting Hawking
radiation. Unfortunately given the rather small values of $M$ that
yield a numerically accessible black hole endpoint, we cannot yet
confirm this behavior. One can certainly see the expected scaling
with $M$ of the stress tensor near the horizon $|T_{uu}|\sim1/M^{4}$,
so we expect future simulations with larger values of $M$ will see
something close to the adiabatic evolution.

\section{Thunderbolt Singularity\label{sec:Thunderbolt-Singularity}}

\begin{figure}

\includegraphics[width=7cm]{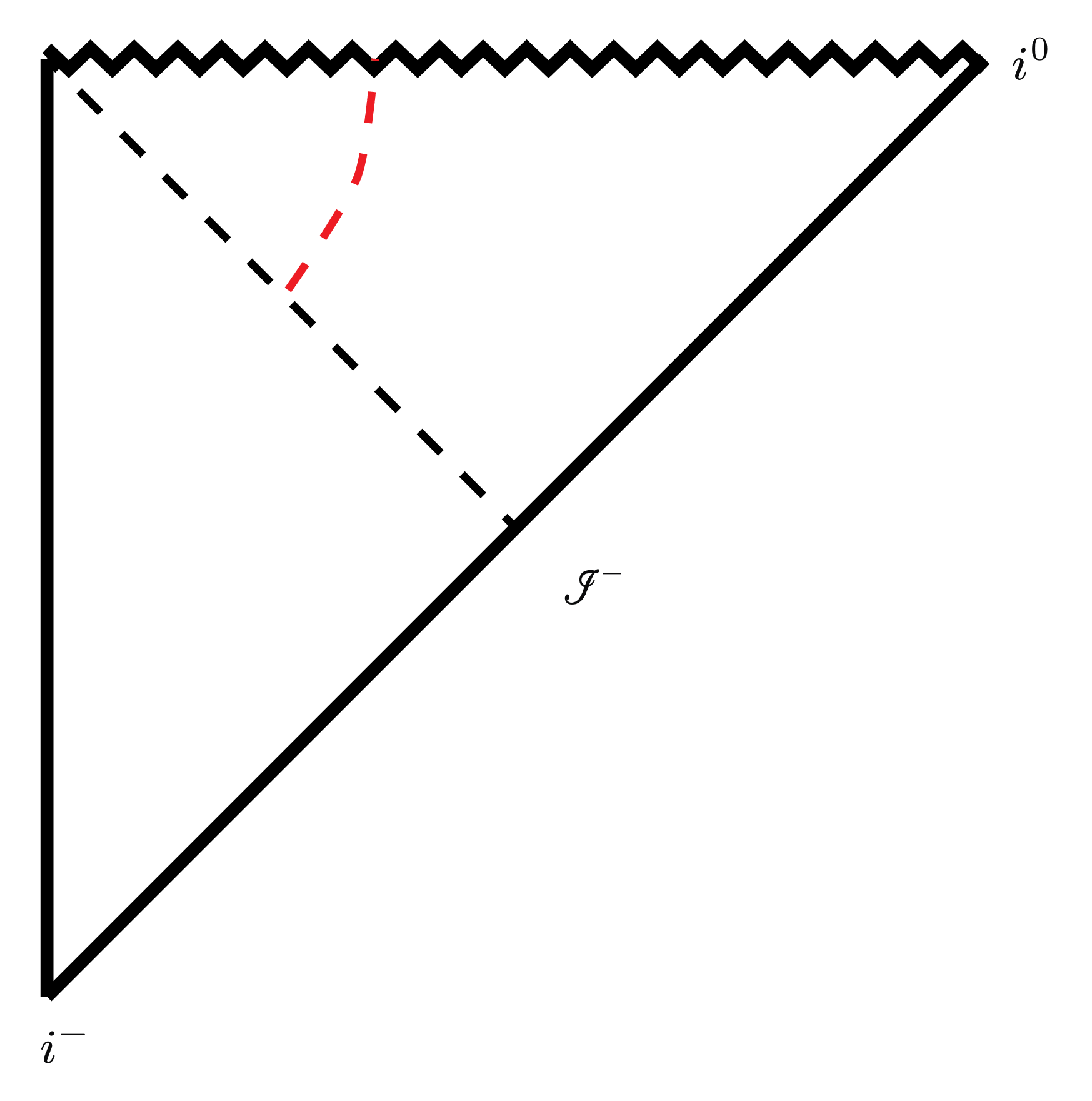}\caption{Penrose diagram for a numerical semiclassical black hole with a spacelike
thunderbolt singularity. The red dashed curve indicates the apparent
horizon of the black hole.\label{fig:Penrose-diagram-for}}
\end{figure}

In this section, we present numerical and analytic evidence for the
emergence of a spacelike \textquotedblleft thunderbolt\textquotedblright{}
singularity as the endpoint of black hole evaporation, as originally
suggested in \citep{Hawking:1992ti}. We show that this feature arises
as a nonlinear instability of the semiclassical equations of motion
and is robust under changes in initial data and numerical resolution.
Remarkably, the singularity develops in regions where the classical
curvature is small, indicating that it is driven by long-range back-reaction
effects rather than strong gravity near the horizon. The global structure
of the spacetime is well described by the Penrose diagram shown in
figure \ref{fig:Penrose-diagram-for}. The singularity appears after
the apparent horizon has receded and extends outward along a spacelike
trajectory.

\begin{figure}

\includegraphics[scale=0.5]{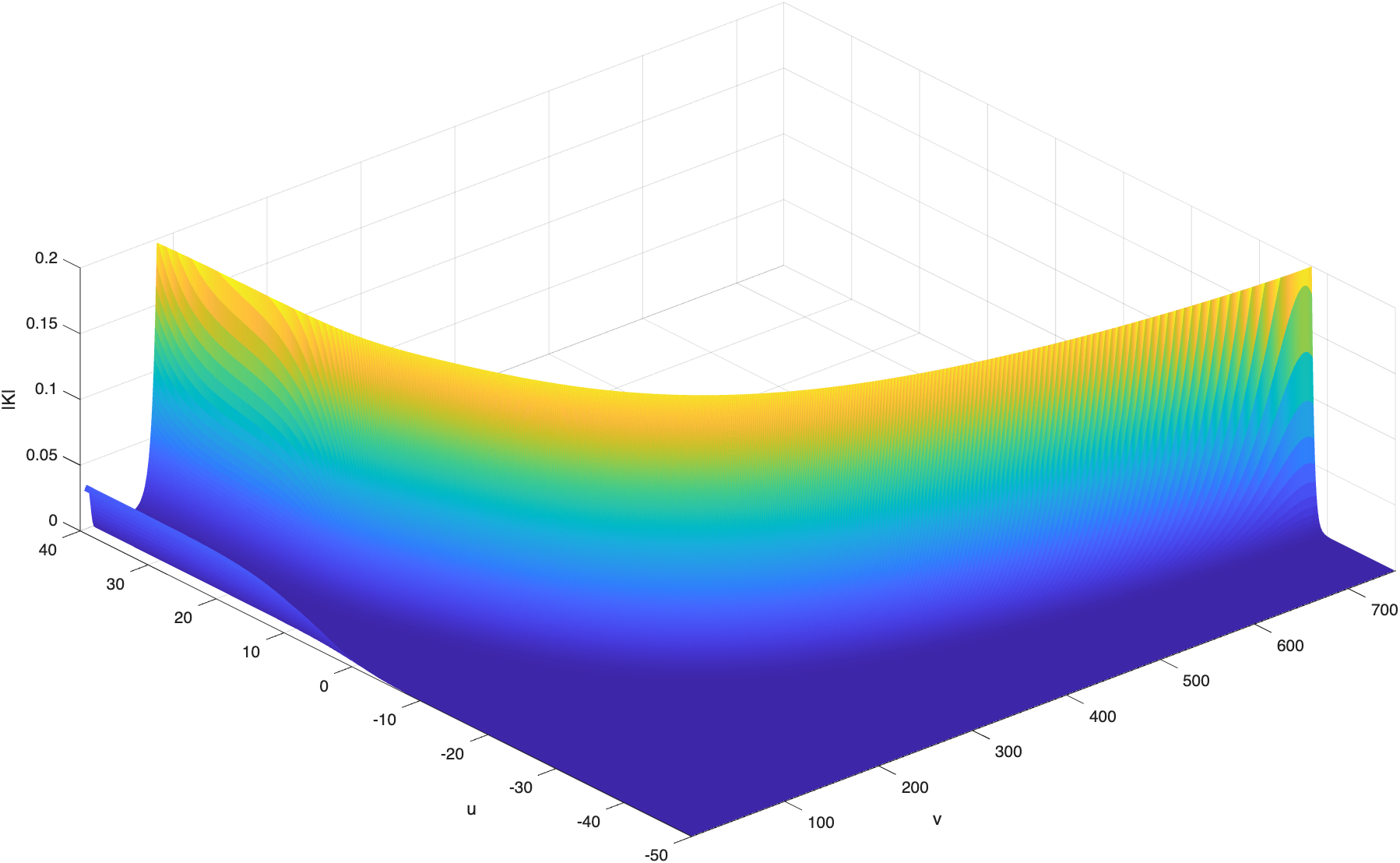}\caption{Absolute value of the Kretschmann scalar for a black hole formed by
an ingoing null shockwave at $v=0$.\label{fig:Kretschmann-scalar-at}}

\end{figure}

To probe this behavior, we perform simulations with $M=2.2$, $A=0$
and anomaly coefficients $a=b=c=10$, evolving the system out to large
values of $v$. The Kretschmann scalar, shown in figure \ref{fig:Kretschmann-scalar-at}
, exhibits a rapid growth signaling the onset of a curvature singularity.
The location of this divergence is consistent with a spacelike surface
extending toward spacelike infinity, in agreement with the Penrose
diagram.

We have tested the robustness of this result by varying the position
of the initial data surface (increasing $|u_{min}|$) and by refining
the numerical grid. This is discussed in more detail in appendix \ref{sec:Numerical-Convergence}. 

To gain analytic insight, we examine the first time step along the
initial slice $u=u_{min}$, subject to the initial conditions \eqref{eq:initialdata}
and \eqref{eq:shockdata}. The equation for $\partial^{2}_{u}\phi$
is linear and can be solved exactly\begin{dmath}
\begin{equation}
\partial^{2}_{u}\phi=\frac{(a+b)(2b-3c)}{384\pi bc}\left(\frac{4}{\left(W\left(e^{\frac{1}{4}(-u_{min}-3)}\right)+1\right)^{3}\left(W\left(e^{\frac{1}{4}(-u_{min}+v-4)}\right)+1\right)}-\frac{1}{\left(W\left(e^{\frac{1}{4}(-u_{min}+v-4)}\right)+1\right)^{4}}-\frac{3}{\left(W\left(e^{\frac{1}{4}(-u_{min}-3)}\right)+1\right)^{4}}\right)\,.\label{eq:phiaccel}
\end{equation}
\end{dmath}This has roughly a linear growth with $v>1$ with a slope
that becomes small as $u_{min}\to-\infty$.\footnote{For the special case $c=2b/3$ the acceleration \eqref{eq:phiaccel}
vanishes on the initial slice. However the solution for the full PDE
is nevertheless qualitatively similar to the numerical solution we
have already discussed.}

Similar behavior is observed in the equations for the $r$ and $\rho$
equations where the corrections to $\partial^{2}_{u}X$ contain slowly
growing terms sourced by the Kretschmann scalar. However, this linear
growth alone is insufficient to produce a singularity. Instead, it
seeds a nonlinear instability that becomes dominant after evolution
in the $u$-direction. Thus, the thunderbolt should be understood
as a genuinely nonlinear phenomenon arising from the higher-derivative
structure of the semiclassical equations. The initial linear growth
provides a perturbation that is amplified by nonlinear interactions,
leading to rapid divergence on a much shorter timescale.

An important question is whether the instability destroys the entire
asymptotic region or remains compatible with a well-defined spacetime
structure. Our numerical results indicate that, for finite $u_{min}$,
the thunderbolt appears at sufficiently large $v$ to be consistent
with the Penrose diagram in figure \ref{fig:Penrose-diagram-for}
. As $|u_{min}|$ is increased, the location of the singularity shifts
outward without approaching a fixed value of $v$, suggesting that
it asymptotes toward spacelike infinity $i^{0}$. We find no evidence
that the thunderbolt intersects future null infinity $\mathscr{I}^{+}$.
Instead, the numerical data are consistent with a spacelike singularity
that caps off the future development of the spacetime while remaining
compatible with causal evolution from regular initial data. 

\begin{figure}
\subfloat[]{\includegraphics[scale=0.3]{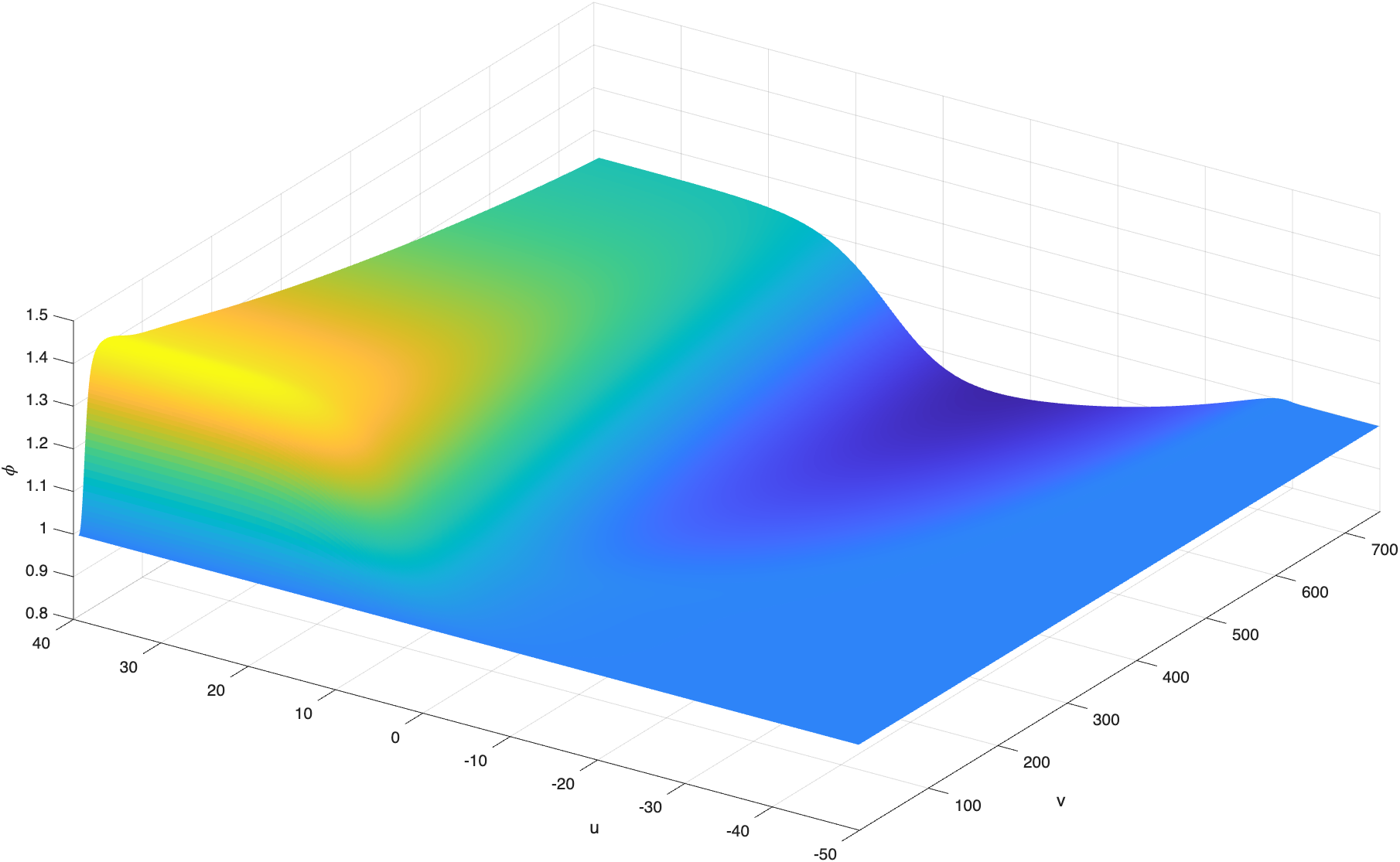}}\\
\subfloat[]{\includegraphics[scale=0.3]{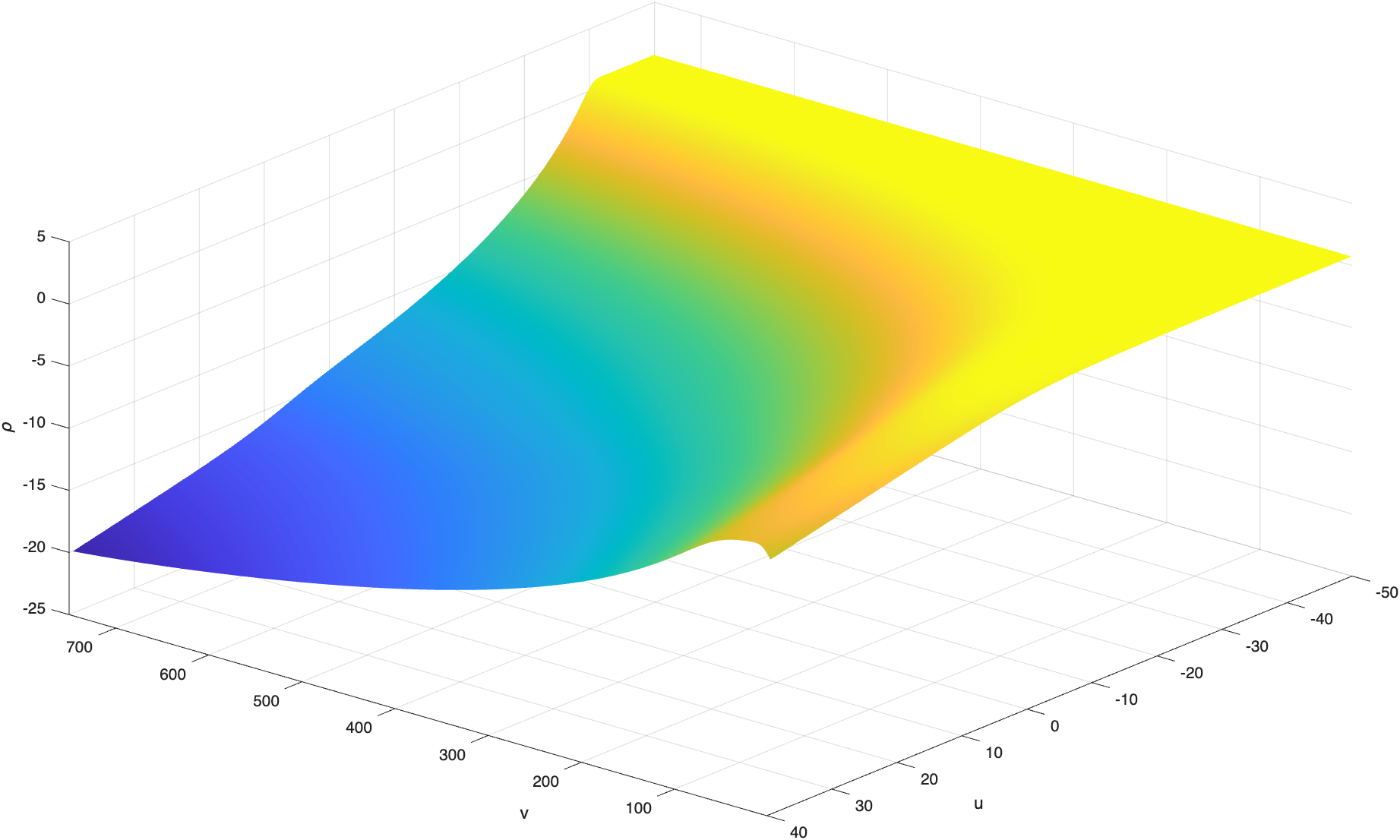}

}\\
\subfloat[]{\includegraphics[scale=0.3]{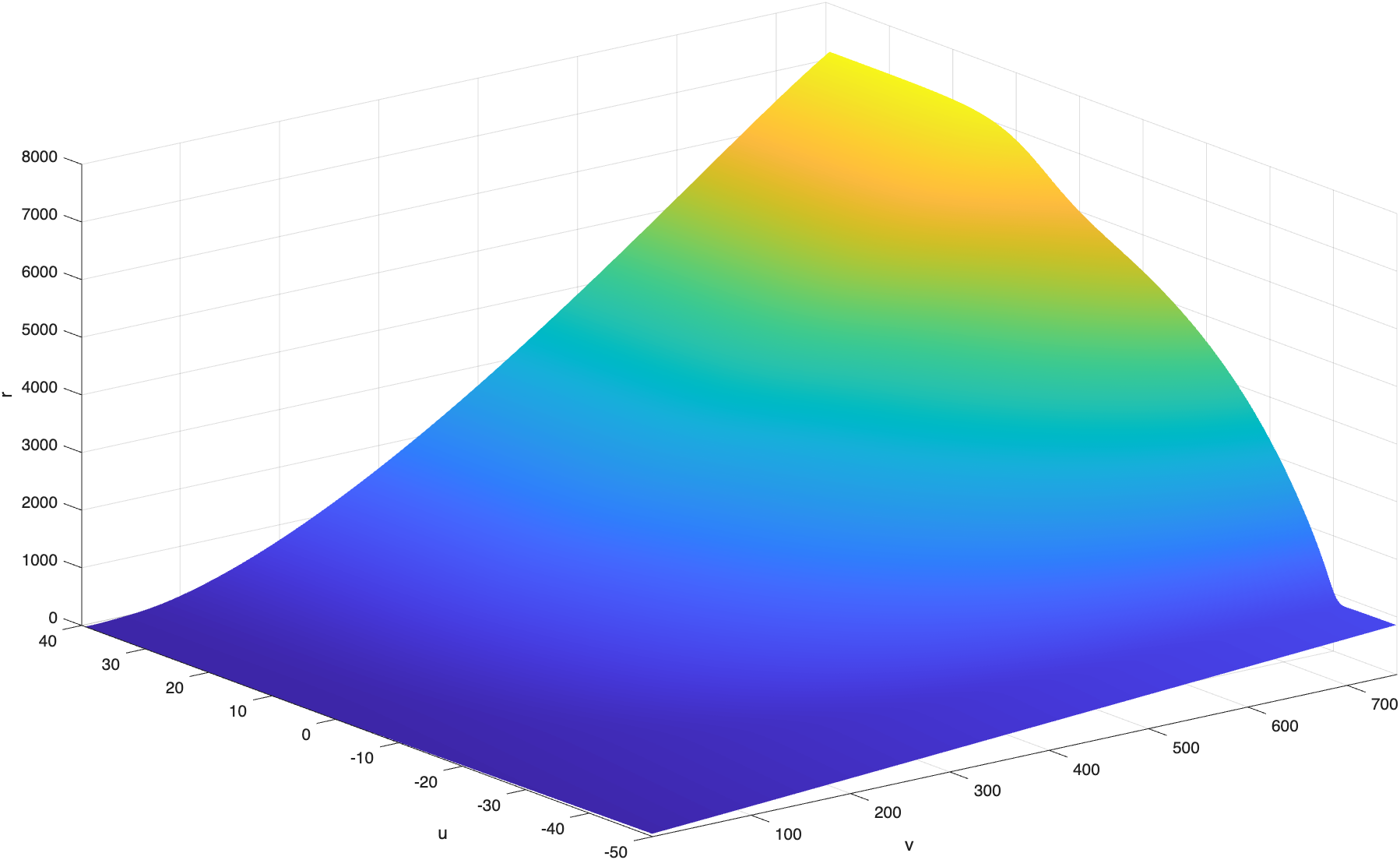}

}\caption{Panels (a),(b) and (c) show surface plots of the auxiliary scalar
$\phi$ and the metric functions $\rho$ and $r$ respectively as
functions of the null coordinates $u$ and $v$. \label{fig:Panels-(a).(b)-and}}

\end{figure}

Further insight into the nature of the instability can be obtained
by examining the behavior of the fields near the singularity. Surface
plots of $\phi$, $\rho$ and $r$ shown in figure \ref{fig:Panels-(a).(b)-and},
reveal a remarkably simple and universal structure. As the thunderbolt
is approached, we observe that:
\begin{itemize}
\item $e^{2\rho}\to0$
\item $\phi\to\mathrm{constant}$
\item $r$ increases as $u$ increases
\end{itemize}
This behavior suggests that the singularity acts as an attractor solution
of the semiclassical equations. The smooth and universal nature of
these profiles provides additional evidence that the thunderbolt is
a genuine physical feature rather than a numerical artifact.

An important question concerns the timescale over which the instability
develops. Due to numerical limitations (i.e. round-off error associated
with double precision arithmetic), we have primarily studied relatively
small black hole masses. However, preliminary results indicate that
the thunderbolt emerges at a roughly fixed interval $\delta v\sim10$
even as the mass is increased (with other parameters held fixed).
If this scaling persists for larger masses, it would imply that the
semiclassical approximation breaks down on a timescale of order the
light-crossing time $M$ rather than the much longer evaporation time
$M^{3}$. This represents a significant departure from the standard
picture of black hole evaporation and suggests that semiclassical
gravity may fail much earlier than previously expected.

It is instructive to compare this behavior with flat spacetime ($M=0$)
where the theory is linearly stable \citep{RIEGERT198456}. In that
case, appropriate boundary conditions are sufficient to eliminate
growing modes. However, in the presence of a black hole background,
nonlinear effects qualitatively change the dynamics, leading to the
instability described above. This is not entirely unexpected: even
in classical general relativity, the proof of nonlinear stability
of Minkowski space is highly nontrivial \citep{Christodoulou}. The
presence of higher-derivative terms in the semiclassical equations
raises the possibility of qualitatively new instabilities.

It is worth noting causality is not violated, despite the apparent
appearance of the spacelike ``faster than light'' thunderbolt emerging
from the black hole endpoint. There exists a well-defined region of
spacetime where the semiclassical equations provide a predictive description
based on regular initial data. However this region of causal evolution
is entirely capped off from future null infinity. The emergence of
the thunderbolt therefore signals a breakdown of the semiclassical
approximation. Rather than representing a physical singularity in
a complete theory of quantum gravity, it should be viewed as an indication
that the effective field theory has reached the limits of its validity
and new quantum effects must be included to correctly describe physics
even far from the black hole.

\section{Conclusions\label{sec:Conclusions}}

We have studied black hole formation and evaporation in a four-dimensional
semiclassical model that preserves diffeomorphism invariance and reproduces
the one-loop trace anomaly. It may be viewed as a truncation of the
exact one-loop effective action where not all conformally invariant
terms are included. By numerically solving the quantum-corrected equations
of motion, we have been able to follow the full evolution of the spacetime,
including the evaporation endpoint and the region beyond.

Our central result is the emergence of a spacelike \textquotedblleft thunderbolt\textquotedblright{}
singularity that develops after the apparent horizon has receded.
This singularity extends outward into regions that would classically
be nearly flat, driven by the back-reaction of the outgoing Hawking
radiation. Both numerical evidence and analytic considerations indicate
that this behavior is a generic feature of four-dimensional semiclassical
gravity with anomaly-induced corrections.

It is conceivable that missing semiclassical corrections corrections
provide a fine-tuning of the initial data that selects among the full
set of semiclassical solutions precisely those who avoid the thunderbolt
instability, as suggested in \citep{Lowe:2022tun}. For example, one
might include a correction to the initial data at order $\hslash$
which induces a flux across the initial slice of order $\hslash^{2}$,
and try to fine tune these corrections to eliminate the thunderbolt.
Testing this hypothesis would require rather delicate numerics that
our current setup cannot provide. Another possibility, which is perhaps
more likely, is that effective field theory requires novel constraints/extensions
to produce thunderbolt free solutions. 

A striking consequence of a thunderbolt singularity is that the breakdown
of the semiclassical approximation does not occur only in regions
of strong curvature near the horizon, but instead extends to large
spacelike separations, including regions that would ordinarily be
considered part of the asymptotic regime. In this sense, semiclassical
effective field theory predicts its own failure over macroscopic distances.

This observation has significant implications for the black hole information
paradox. The standard formulation assumes that semiclassical effective
field theory remains valid arbitrarily far from the black hole, allowing
one to track entanglement between outgoing Hawking radiation and interior
degrees of freedom. The emergence of the thunderbolt invalidates this
assumption: the semiclassical description breaks down before such
an analysis can be consistently completed.

Any complete theory of quantum gravity that resolves the thunderbolt
must therefore modify the semiclassical picture at large distances.
In particular, it must alter the structure of quantum correlations
in the Hawking radiation, potentially providing a mechanism by which
information is preserved without requiring violations of causality
or locality in the usual sense. From this perspective, the thunderbolt
should not be viewed as a physical singularity that would occur in
nature, but rather as a diagnostic signal indicating the limits of
semiclassical reasoning. Its presence identifies a regime in which
new physics must enter and constrains the way in which a consistent
quantum theory of gravity can resolve the evaporation process.

Finally, our results suggest that near the black hole the breakdown
of semiclassical gravity may occur on timescales much shorter than
the naive evaporation time, potentially of order the light-crossing
time for sufficiently large black holes. This raises the tantalizing
possibility that observable deviations from semiclassical expectations
could arise in realistic astrophysical settings.

Understanding how a complete quantum theory of gravity eliminates
or resolves the thunderbolt instability remains an important open
problem for future work. Progress in this direction may provide crucial
insight into the fate of evaporating black holes and the ultimate
resolution of the information paradox.

\appendix

\section{Numerical Convergence\label{sec:Numerical-Convergence} }

It is important to check that the thunderbolt instability is a genuine
physical effect and not simply an artifact of the numerical method.
In order to test this, we demonstrate pointwise convergence of the
solution for $\rho$ in the neighborhood of the onset of the thunderbolt.
A distinctive signature of the onset of the thunderbolt is that the
metric field $\rho$ turns over when traced along a light-like line
at constant $u$, leading to $e^{2\rho}\to0$. One would have ordinarily
expected that $\rho$ approaches a constant as a light ray moves out
into the asymptotically flat region. We consider simulations with
$M=2.2$ and stretching parameter $A=0.15$ and focus on a light-ray
that emerges from the apparent horizon. We show the pointwise convergence
of $\rho$ by plotting various grid sizes and values of $u_{min}$
on the same plot. This results agree to within less that $1\%$ demonstrating
numerical convergence. This behavior is qualitatively the same for
any outgoing null ray outside the apparent horizon.
\begin{figure}

\includegraphics[scale=0.4]{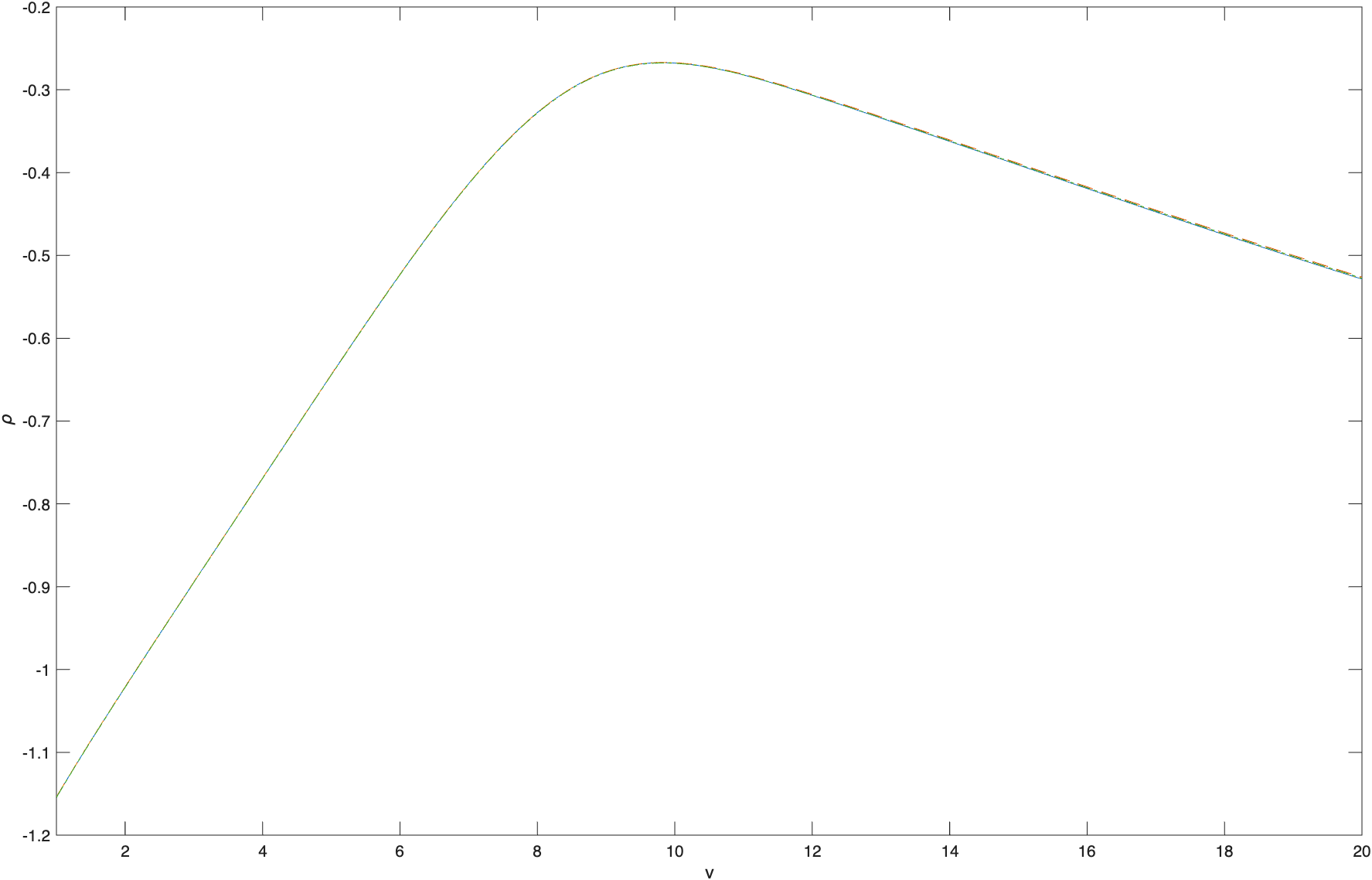}\captionsetup{justification=raggedright,singlelinecheck=false}\caption{The value of $\rho$ is shown along a line of constant $u$ which
emerges from behind the apparent horizon. Here $u=13$. The results
are shown for $u_{min}=-50$ and $u_{min}=-100$ and for grid sizes
$20k\times2k$ and $10k\times1k$, using solid, dashed, dotted and
dot-dashed lines respectively, which can barely be distinguished on
the right side of the plot. The lines coincide well past the peak
which signals the onset of the thunderbolt instability.}

\end{figure}

\bibliographystyle{utcaps}
\bibliography{riegert}

\end{document}